# Does the size distribution of mineral dust aerosols depend on the wind speed at emission?

**Jasper F. Kok**[1,*]

[1]Advanced Study Program, National Center for Atmospheric Research, Boulder, CO, USA
[*]Now at Department of Earth and Atmospheric Sciences, Cornell University, Ithaca, NY, USA
Correspondence to: J. F. Kok (jfkok@umich.edu)

**Abstract**

The size distribution of mineral dust aerosols partially determines their interactions with clouds, radiation, ecosystems, and other components of the Earth system. Several theoretical models predict that the dust size distribution depends on the wind speed at emission, with larger wind speeds predicted to produce smaller aerosols. The present study investigates this prediction using a compilation of published measurements of the size-resolved vertical dust flux emitted by eroding soils. Surprisingly, these measurements indicate that the size distribution of naturally emitted dust aerosols is independent of the wind speed. The recently formulated brittle fragmentation theory of dust emission is consistent with this finding, whereas other theoretical dust emission models are not. The independence of the emitted dust size distribution with wind speed simplifies both the interpretation of geological records of dust deposition and the parameterization of dust emission in atmospheric circulation models.

## 1. Introduction

Mineral dust aerosols affect the Earth system through a wide range of interactions, which include scattering and absorbing radiation, serving as cloud condensation and ice nuclei, providing nutrients to ecosystems, and lowering the reflectivity of snow and ice (e.g., Goudie and Middleton, 2006; Mahowald et al., 2010; DeMott et al., 2010; Painter et al., 2010). The size of dust aerosols affects many of these interactions and also determines the lifetime and thus transport of dust (e.g., Tegen and Lacis, 1996). Moreover, the deposition of dust aerosols into deep sea sediments (Rea, 1994), ice cores (Ruth et al., 2003), and loess deposits (Ding et al., 2002) provides important information about past climate regimes. For these reasons, a detailed understanding of the particle size distribution (PSD) is critical to improving our understanding of the myriad interactions between mineral dust aerosols and the Earth system.

However, large uncertainties exist in the treatment of the emitted dust PSD in atmospheric circulation models (Cakmur et al., 2006; Kok, 2011). In particular, it is unclear whether the emitted dust PSD depends on the wind speed at emission (Sow et al., 2009; Shao et al., 2011). Determining this dependence will thus facilitate more accurate simulations of dust interactions with weather, climate, and ecosystems, as well as aid the interpretation of variations in the mean diameter of deposited dust in geological records (Rea, 1994; Ding et al., 2002; Ruth et al., 2003).

Measurements of the dependence of the emitted dust PSD on wind speed have yielded contradictory results. Whereas a subset of wind tunnel studies have reported that the dust aerosol size decreases with increasing wind speed (Alfaro et al., 1997, 1998; Alfaro, 2008), other wind tunnel measurements and field studies have not found a clear dependence of the emitted dust PSD on the wind speed (Gillette et al., 1974; Shao et al., 2011). Theoretical models of dust emission mirror these contradictory experimental results: whereas the models of both Shao (2001, 2004) and Alfaro and Gomes (2001) predict that the size of emitted dust aerosols decreases with wind speed, the recently formulated brittle fragmentation theory of dust emission predicts that the emitted dust PSD is independent of the wind speed (Kok, 2011). While that former theories are in agreement with a subset of wind tunnel studies, the latter theory is in agreement with field measurements (Fig. 1).

In order to (i) help distinguish between these contrasting theoretical dust emission models, (ii) inform dust emission parameterizations in atmospheric circulation models, and (iii) aid the interpretation of geological dust deposition records, this article for the first time uses a compilation of published measurements to determine the dependence of the emitted dust PSD on the wind speed. As shown in the subsequent sections, the results indicate that the emitted dust PSD is invariant to even substantial changes in wind speed.

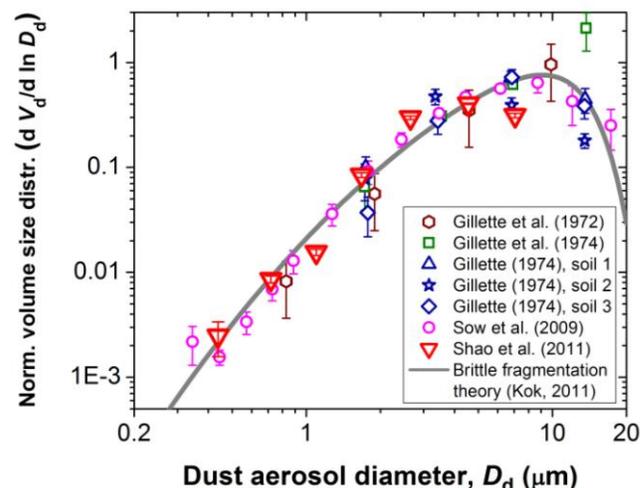

Figure 1. Field measurements with standard error of the volume size distribution of emitted dust aerosols (assorted symbols), processed as described in Kok (2011). The brittle fragmentation theory of dust emission (solid line) (Kok, 2011) is in good agreement with these measurements, including the subsequently published Shao et al. (2011) (large triangles; note that Shao et al. (2011) questioned the reliability of their 0.3 – 0.6 μm particle size bin).

Table 1. Summary of published wind tunnel and field measurements of the size-resolved vertical dust flux. Listed for each data set are the number of measurements, the friction speed and dust aerosol diameter ranges spanned by those measurements, the average $D_N$ and $D_V$ with standard deviation of those measurements, and the trend with standard error of $D_N$ and $D_V$ with $u_*$ for each data set. This information is also provided for a compilation of all six data sets of field measurements.

| Data set | Measurement type | Number of measurements | $u_*$ range (m/s) | $D$ range (μm) | Average $D_N$ (μm) | $D_N$ trend (μm m$^{-1}$ sec) | Average $D_V$ (μm) | $D_V$ trend (μm m$^{-1}$ sec) |
|---|---|---|---|---|---|---|---|---|
| Gillette et al. (1974) | Wind tunnel | 3 | 0.74 – 1.14 | 1.2 – 80 | 2.6 ± 0.2 | 0.1 ± 0.7 | 5.0 ± 0.5 | 2.9 ± 1.9 |
| Alfaro et al. (1998) | Wind tunnel | 4 | 0.35 – 0.66 | 1 – 100 | 4.6 ± 2.7 | -21.2 ± 0.1 | 6.0 ± 1.4 | -11.1 ± 0.1 |
| Gillette et al. (1974) | Field | 3 | 0.25 – 0.78 | 1.2 – 40 | 2.6 ± 0.5 | -1.0 ± 1.0 | 5.4 ± 0.3 | -0.1 ± 2.0 |
| Gillette (1974), soil 1 | Field | 12 | 0.18 – 0.58 | 1.2 – 20 | 2.5 ± 0.4 | -2.6 ± 1.6 | 5.2 ± 0.7 | -2.6 ± 3.0 |
| Gillette (1974), soil 2 | Field | 4 | 0.49 – 0.78 | 1.2 – 20 | 2.6 ± 0.2 | 1.7 ± 2.0 | 4.5 ± 0.5 | 3.6 ± 5.1 |
| Gillette (1974), soil 3 | Field | 4 | 0.28 – 0.48 | 1.2 – 20 | 3.1 ± 0.4 | 3.8 ± 4.3 | 5.4 ± 0.7 | 3.7 ± 6.1 |
| Sow et al. (2009) | Field | 3 | 0.40 – 0.60 | 0.3 – 20 | 2.6 ± 0.2 | -1.6 ± 2.9 | 5.0 ± 0.3 | -2.9 ± 5.8 |
| Shao et al. (2011) | Field | 8 | 0.20 – 0.55 | 0.6 – 8.4[a] | 2.3 ± 0.1 | 0.3 ± 0.6 | 4.4 ± 0.2 | 0.7 ± 1.4 |
| All field measurements | Compilation | 34 | 0.18 – 0.78 | 1.2 – 8.4 | 2.6 ± 0.4 | 0.1 ± 0.4 | 4.9 ± 0.6 | 0.5 ± 0.8 |

[a] The Shao et al. measurements actually span the size range of 0.3 – 8.4 μm, but the authors questioned the reliability of the 0.3 – 0.6 μm size bin (Shao et al., 2011, p. 13).

## 2. Methods

To investigate whether the dust PSD depends on the wind speed at emission, I determine the variation of the mean dust aerosol diameter with the wind friction speed $u_*$ (defined as the square root of the ratio of the wind stress and the air density). I do so by calculating the mean aerosol diameters by number ($D_N$) and volume ($D_V$) for every reported value of $u_*$ of each published data set of the size-resolved vertical dust flux emitted by an eroding soil (see Table 1 and Section 2.1). That is,

$$D_N = \int_{D_{low}}^{D_{up}} D \frac{dN}{dD} dD \Big/ \int_{D_{low}}^{D_{up}} \frac{dN}{dD} dD$$
$$D_V = \int_{D_{low}}^{D_{up}} D \frac{dV}{dD} dD \Big/ \int_{D_{low}}^{D_{up}} \frac{dV}{dD} dD \quad (1)$$

where $N$ and $V$ respectively denote the number and volume of emitted aerosols of a given diameter $D$. The integration limits $D_{low}$ = 1.2 μm and $D_{up}$ = 8.4 μm maximize the overlap in size ranges measured by the various data sets (Table 1).

The detailed procedure for using Eq. (1) to calculate $D_N$, $D_V$, and their uncertainties is described in the supplementary text. Briefly, data sets for which the particle bin limits do not exactly match $D_{low}$ or $D_{up}$ were corrected by truncating the relevant particle bin(s). Furthermore, the integration in Eq. (1) was performed by assuming that the sub-bin distribution follows the power law for ~2–10 μm diameter dust reported in Gillette et al. (1974) and Kok (2011) (i.e., $dN/d\log D \sim D^{-2}$ and $dV/d\log D \sim D$). Finally, the uncertainties of $D_N$ and $D_V$ were calculated by propagating the uncertainty in the measurements of $N(D)$ and $V(D)$, respectively.

### 2.1 Description of data sets used

Determining the size-resolved vertical dust flux emitted by an eroding soil requires simultaneous measurements of the wind speed and the size-resolved dust aerosol concentration for at least two separate heights (Gillette et al., 1972). Since these measurements are difficult to make, only a limited number of data sets exist in the literature.

The first field measurements of the size-resolved vertical dust flux were made by Gillette and co-workers. Specifically, Gillette (1974) and Gillette et al. (1974) reported measurements of two fine sand soils and two loamy fine sand soils in Texas for wind friction speeds of 0.18–0.78 m/s. These measurements were made using two single-stage jet impactors at heights of 1.5 and 6 meters. The collected aerosols were subsequently analyzed using microscopy to retrieve the size-resolved vertical flux of dust aerosols larger than 1.2 μm in diameter.

More recently, Sow et al. (2009) used two optical particle counters at heights of 2.1 and 6.5 meters to measure the size-resolved vertical flux of dust aerosols larger than 0.3 μm. They reported measurements made during three dust storm events in Niger for which the average wind friction speed varied between 0.4 and 0.6 m/s. Sow et al. (2009) did not report the soil type. Finally, the recent study of Shao et al. (2011) measured the vertical flux of dust aerosols with diameters of 0.3 – 8.4 μm using three optical particle counters at 1.0, 2.0, and 3.5 meters above a loamy sand agricultural soil in Australia (Ishizuka et al., 2008). Shao et al. (2011) reported measurements for wind friction speeds in bins ranging from < 0.20 m/s to 0.55 m/s.

In addition to these field measurements, several wind tunnel studies of the size-resolved vertical dust flux have been performed. The first of these was reported by Gillette et al. (1974) for friction speeds between 0.74 and 1.14 m/s for the same soil as used in field measurements. Subsequent wind tunnel measurements by Alfaro et al. (1998) used a Spanish loamy soil and reported the size distribution of emitted dust aerosols with diameters between 1 and 100 μm for wind friction speeds between 0.35 and 0.66 m/s.

All data sets of the size-resolved vertical dust flux used in this study are summarized in Table 1.

## 3. Results

Results from a compilation of the six data sets of field measurements (see Table 1) show that the trends of $D_N$ and $D_V$ with $u_*$ are within the standard error, and thus statistically insignificant (Figs. 2a, b and Table 1). Similarly, individual field data sets show opposing and mostly statistically insignificant trends of $D_N$ and $D_V$ with $u_*$ (Table 1). In addition to this apparent insensitivity to $u_*$, $D_N$ and $D_V$ also seem relatively insensitive to changes in the soil characteristics. Indeed, the mean aerosol diameters of different field data sets are mostly within one standard deviation (Table 1), despite variations in soil characteristics between the different data sets (see Section 2.1). Note that calculating $D_N$ and $D_V$ with the extended size range of $D_{low} = 0.6$ μm and $D_{up} = 8.4$ μm spanned by the most recent field measurements of Sow et al. (2009) and Shao et al. (2011) yields qualitatively similar results (Supplementary Figure S1).

Although field measurements thus indicate that the PSD of naturally emitted dust does not depend on $u_*$, the wind tunnel measurements of Alfaro et al. (1998) do show a statistically significant decrease of the mean aerosol diameter with $u_*$ (Fig. 2a, b and Table 1). Comparable wind tunnel measurements by Alfaro et al. (1997) and Alfaro (2008) were not included in the present analysis because these studies did not use natural soil and measured the emitted dust PSD only up to 5 μm, respectively. However, these measurements similarly show a pronounced shift to smaller aerosol diameters with $u_*$ (see figure 10 in Alfaro et al. (1997) and figure 2c in Alfaro (2008)). In contrast to these results by Alfaro and colleagues, the wind tunnel measurements of Gillette et al. (1974) show no clear dependence of $D_N$ and $D_V$ on $u_*$ (Fig. 2a, b and Table 1).

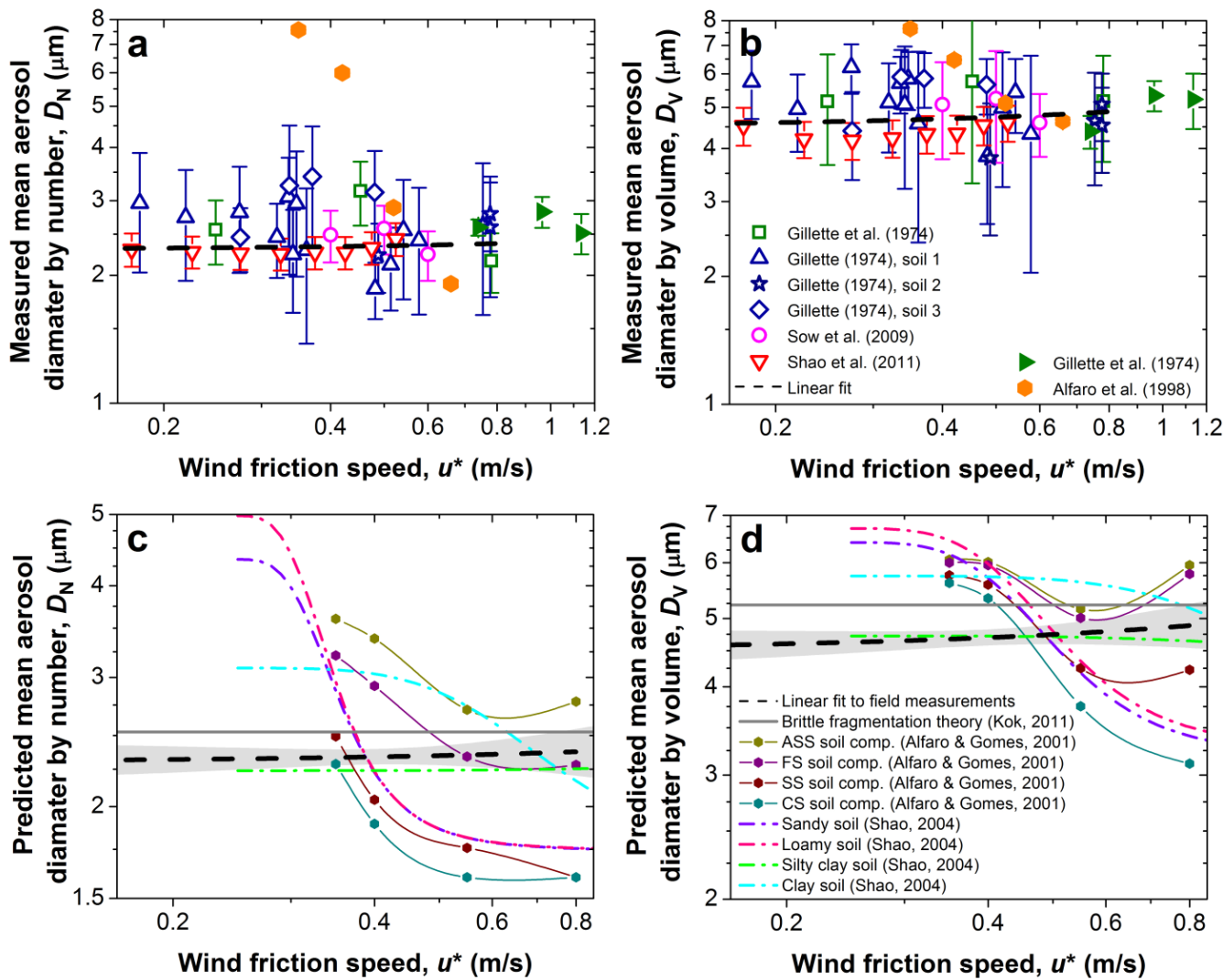

Figure 2. Dependence of the mean dust aerosol diameters by number (**a**) and volume (**b**) on the wind friction speed. Wind tunnel and field measurements are respectively denoted by filled and open symbols. Linear least-squares fits to the compilation of all field measurements (black dashed lines) show trends of the mean aerosol diameter with $u_*$ that are within one standard error (Table 1), and are thus statistically insignificant. Linear fits to individual data sets are also reported in Table 1. Panels (**c**) and (**d**) respectively show the mean aerosol diameter by number and volume predicted by theoretical dust emission models. Plotted for comparison are the linear fits to the measurements from (a) and (b); the shading denotes the uncertainty on the fit, which is calculated as described in the supplementary text. Parameters required for the brittle fragmentation theory (solid grey line) were obtained from Kok (2011). The models of Alfaro and Gomes (2001) (solid lines and hexagons) and Shao (2004) (dash-dotted lines) require detailed soil size distribution information, which is not available for most of the data sets of the size-resolved vertical dust flux. Nonetheless, mean dust aerosol diameters were calculated from Eq. (1) for several 'typical' arid soils, which thus do not necessarily correspond to any of the soils for which measurements of the emitted dust PSD were made. For Alfaro and Gomes (2001), $D_N$ and $D_V$ were calculated from the theoretical emitted dust PSD reported for four values

of $u_*$ in their table 5. The increase in $D_N$ and $D_V$ at $u_* = 0.80$ m/s for several of the soils is inconsistent with the assumption in Alfaro and Gomes (2001) that higher wind speeds produce more disaggregated aerosols and might be due to numerical errors in the production of their table 5 (Grini et al., 2002). For Shao (2004), $D_N$ and $D_V$ were obtained by inserting the four soil size distributions reported in his table 1 into his Eq. 6, and using a threshold $u_*$ for erosion of 0.25 m/s, consistent with the thresholds reported in the experimental data sets used here (Table 1).

## 4. Discussion
### 4.1 Testing the accuracy of theoretical dust emission models

The surprising result that the PSD of naturally emitted dust aerosols does not depend on $u_*$ can be used to test the accuracy of theoretical dust emission models. The brittle fragmentation theory of dust emission (Kok, 2011) correctly predicts this independence (Figs. 2c, d), whereas the dust emission theories of Alfaro and Gomes (2001) and Shao (2001, 2004) predict that larger wind speeds produce more disaggregated and hence smaller dust aerosols. These theories thus predict a decrease of the mean aerosol diameters with increasing $u_*$ (Figs. 2c, d). However, this prediction is inconsistent with measurements (Figs. 2a, b), except for the Alfaro et al. (1997, 1998) wind tunnel studies. A possible explanation for the puzzling discrepancy of the Alfaro et al. experiments with other measurements is discussed in Section 4.2.

The results of the detailed field study of Sow et al. (2009) further favor the brittle fragmentation theory of dust emission. Although Sow et al. (2009) found that the emitted dust PSD is invariant to variations in $u_*$ during a given dust event, consistent with the findings in Table 1 and Fig. 2, they did find variations in the emitted dust PSD between dust events (see their figures 10 and 9, respectively). Moreover, Sow et al. reported changes in the aerodynamic roughness length (table 1 in Sow et al. (2009)) and the threshold $u_*$ for dust emission between the three measured dust events, which could indicate changes in the physical state of the soil. Brittle fragmentation theory (Astrom, 2006) predicts that changes in the physical state of the brittle material (aggregates of dust particles in the soil in this case) affect the propagation distance λ of the side branches of cracks created by a fragmenting impact. For instance, precipitation between the dust emission events could have affected the cohesiveness of the soil dust aggregates (Rice et al., 1996) and thus changed the propagation distance λ. Although these changes affect the large-size cutoff, which is determined by λ and is on the order of 10 – 15 μm, they do not affect the emitted dust PSD in the <~5 μm size range, which is instead determined by the fully dispersed (and presumably constant) soil PSD (Kok, 2011). This 'fingerprint' of brittle fragmentation theory is indeed apparent in the measurements of Sow et al. (2009), which are highly similar between the three dust emission events for the < 5 μm size range, yet show variation in the > 5 μm size range (see figure 9 in Sow et al. (2009)). By adjusting the value of λ, this variation of the emitted dust PSD between the three events is reproduced by brittle fragmentation theory (Fig. 3).

### 4.2 Cause of discrepancy between theories and measurements

As discussed above, the theories of Alfaro and Gomes (2001) and Shao (2001, 2004) predict a shift to smaller aerosol diameters with increasing $u_*$ (see figure 5 in both Shao (2001) and Alfaro and Gomes (2001)), which field measurements indicate is incorrect (Table 1 and Figs. 2a, b). This predicted shift is due to the assumption in these models that the energy with which bouncing ("saltating") sand particles impact the soil is proportional to $u_*^2$ (Eq. 1 in Alfaro and Gomes (2001) and p. 20,247 in Shao (2001)). (The breakdown of dust aggregates by the impact of these saltating particles on the soil is the main source of dust aerosols (e.g., Gillette et al., 1974).) The Alfaro and Gomes (2001) and Shao (2001, 2004) models then hypothesize that the increase of the saltator impact energy with $u_*$ produces more disaggregated and hence smaller dust aerosols.

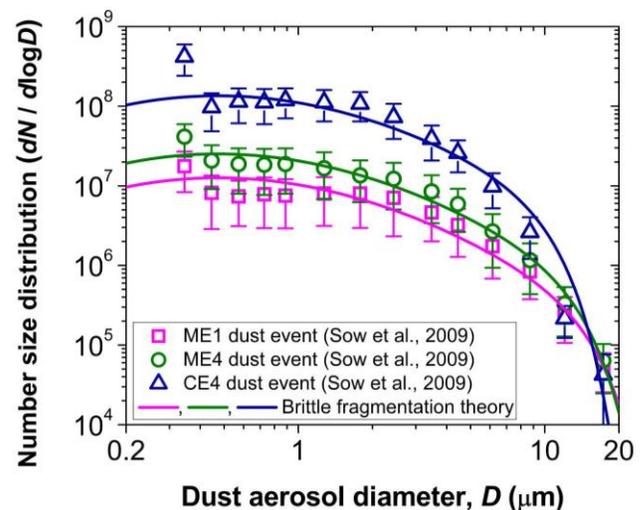

Figure 3. Measurements by Sow et al. (2009) of the emitted dust PSD for three different dust events are reproduced by varying the side crack propagation length λ in the brittle fragmentation theory of dust emission (Kok, 2011). Values for λ of 15.1, 13.5, and 10.3 μm were obtained for respectively the ME1, ME4, and CE4 dust events by using a least-squares fitting procedure with Eq. (5) in Kok (2011). The fully-dispersed soil PSD parameters in the brittle fragmentation theory were obtained from Kok (2011).

Although these arguments appear plausible, recent measurements, numerical models, and theories of saltation all indicate that the saltator impact speed, and thus the impact energy, does not depend on $u_*$ (Fig. 4). This result is a logical consequence of the requirement that exactly one particle must be ejected from the soil bed for each particle impacting it in order for saltation to be in steady state. This condition is fulfilled at a particular mean saltator impact speed that is independent of $u_*$ (Ungar and Haff, 1987; Andreotti, 2004; Kok and Renno, 2009; Kok, 2010a; Duran et al., 2011). Since numerical models and field measurements of saltation indicate that the saltation flux responds to variations in wind speed on a characteristic time scale of a second (Anderson and Haff, 1988; McEwan and Willetts, 1993; Jackson and McCloskey, 1997), saltation in most natural conditions can be considered to be close to steady state (Duran et al., 2011). Removing the assumption that saltator impact speeds increase with $u_*$ would thus likely improve the agreement of the Alfaro and Gomes (2001) and Shao (2001, 2004) theoretical models with field measurements (see Figs. 2c, d), as also inferred by Shao et al. (2011, p. 18) from their field measurements.

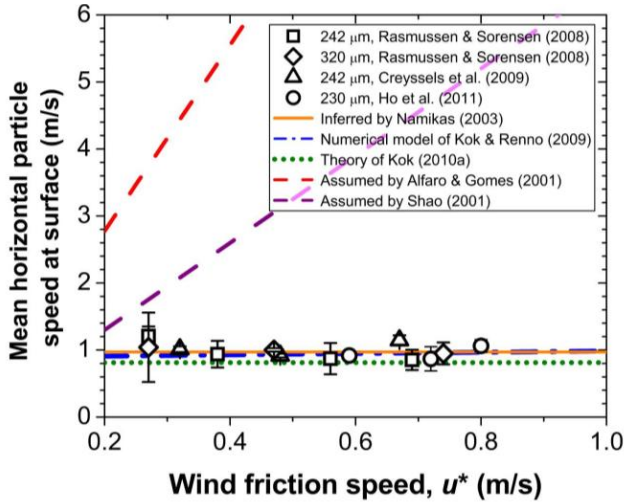

Figure 4. Wind tunnel measurements of the speed of ~250 – 300 μm saltating particles (symbols) indicate that the mean horizontal speed at the surface stays constant with $u_*$. Similarly, Namikas (2003) inferred from his field measurements that the speed with which saltating particles are launched from the surface is independent of $u_*$ (solid orange line). These experimental results are supported by predictions for 250 μm sand by both theory (dotted green line denotes Eqs. 13 and 14 of Kok, 2010a; see also Ungar and Haff, 1987 and Duran et al., 2011) and a recent numerical model (dash-dotted blue line; Kok and Renno, 2009). The assumption of increasing saltator speed by Alfaro and Gomes (2001; dashed red line) and Shao (2001; dashed purple line) is thus likely incorrect. (The impact speed of $v_{imp} = 20u_*$ assumed by Alfaro and Gomes (see their Eq. 1) and the launch speed and angle of ~0.70 m/s and ~35° degrees inferred by Namikas (2003) were converted to a mean horizontal surface speed by using that the rebound speed is ~$v_{imp}$/2, and that the impact and launch angles are ~12° and ~35° degrees (Kok and Renno, 2009). The wind tunnel measurements of Rasmussen and Sorensen (2008), Creyssels et al. (2009), and Ho et al. (2011) were extrapolated to the surface as detailed in Kok (2010b).)

Note that the arguments above apply only to *transport limited* saltation, for which the amount of saltating sand is limited by the availability of wind momentum to transport the sand (Nickling and McKenna Neuman, 2009). Indeed, in the alternative case of *supply limited* saltation, for which the amount of saltating sand is limited by the availability of loose soil particles that can participate in saltation, the drag on the wind by saltating particles is insufficient to reduce the saltator impact speed to its wind-independent value. Consequently, the impact speed in supply limited saltation generally increases with wind speed (Houser and Nickling, 2001; Ho et al., 2011). Since none of the experimental studies of size-resolved dust emissions (Table 1) reported supply limited conditions, it is thus possible that the PSD of dust aerosols generated during supply limited saltation does depend on the wind speed. However, since supply limited saltation occurs due to aggregation of the soil surface or the formation of crusts, such soils usually have a higher threshold wind speed for dust emission. Consequently, supply limited soils are inherently less productive sources of dust aerosols (Rice et al., 1996; Lopez et al., 1998; Gomes et al., 2003; Nickling and McKenna Neuman, 2009).

The arguments above also provide a possible explanation for the puzzling result of the wind tunnel measurements of Alfaro and colleagues (Alfaro et al., 1997, 1998; Alfaro, 2008), which found a strong dependence of the emitted dust aerosol size distribution on $u_*$, in conflict with results from both field measurements and the wind tunnel study of Gillette et al. (1974). The cause of this discrepancy might be that Alfaro and colleagues used a wind tunnel with a working section of only 3.1 meters in length (Alfaro et al., 1997). Indeed, measurements indicate that ~10 meters is required to produce steady-state saltation for natural soils (Shao and Raupach, 1992; Duran et al., 2011), although the use of carefully designed roughness elements can reduce this length (Rasmussen et al., 1996, 2009). Consequently, saltation did probably not reach steady-state in the wind tunnel used by Alfaro et al. (1997), as also noted by these authors (p. 11,243). Therefore, the steady-state requirement that there must be exactly one particle leaving the soil bed for each particle impacting it, which constrains the saltator impact speed to remain constant with $u_*$ (see Fig. 4 and discussion above), did probably not apply. Increases in wind speed could thus have produced increases in the saltator impact speed, which in turn could have produced smaller dust aerosols (Alfaro and Gomes, 2001). This interpretation is supported by the apparent independence of the mean aerosol diameter with $u_*$ for the wind tunnel measurements of Gillette et al. (1974), which were performed in a wind tunnel with a longer working section of 7.2 meters.

## 5. Summary and Conclusions

The present study for the first time uses a compilation of published measurements of the size-resolved vertical dust flux emitted by eroding soils to determine the dependence of the emitted dust PSD on wind speed. The results indicate that the size distribution of naturally emitted dust aerosols is independent of the wind speed at emission (Fig. 2a,b and Table 1). This finding is important for several reasons. First, it simplifies the parameterization of dust emission in atmospheric circulation models, many of which currently account for a dependence of the emitted dust PSD on the wind speed (e.g., Ginoux et al., 2001). Second, this finding simplifies the interpretation of geological records of dust deposition. Indeed, it supports the interpretation that increases in the mean dust size in these records are not related to changes in the wind speed during emission, but instead indicate either stronger transporting winds or a reduced distance to the source (Ding et al., 2002; Ruth et al., 2003). And finally, the independence of the emitted dust PSD with wind speed can be used to test the accuracy of theoretical dust emission models. Specifically, the models of both Alfaro and Gomes (2001) and Shao (2001, 2004) predict that larger wind speeds produce smaller dust aerosols, which is thus inconsistent with measurements. The cause of this discrepancy is probably the assumption by these models that the speed of impacting saltators is proportional to the wind friction speed, which is likely incorrect (Fig. 4). In contrast, the brittle fragmentation theory of dust emission (Kok, 2011) does correctly predict the independence of the emitted dust PSD with wind speed, and is also consistent with the variation of the coarse dust fraction (> ~5 μm) with changes in the soil state observed by Sow et al. (2009) (Fig. 3).

**Acknowledgements**

The author thanks Shanna Shaked, Alexandra Jahn, and Samuel Levis for critical comments on the manuscript, Natalie Mahowald for helpful discussions, and Masahide Ishizuka for providing the raw data from figure 12 of Shao et al. (2011), which was used in Figures 1 and 2 of the present article. The National Center for Atmospheric Research (NCAR) is sponsored by the National Science Foundation.

**Supplement related to this article is available online at:**
http://www.atmos-chem-phys.net/11/10149/2011/
acp-11-10149-2011-supplement.pdf.